\documentclass[runningheads]{llncs}
\usepackage[T1]{fontenc}
\usepackage[english]{babel}
\usepackage{tikz}
\usetikzlibrary{calc,positioning,fit}
\usetikzlibrary{shapes}
\usetikzlibrary{shapes.arrows}
\usetikzlibrary{arrows.meta}
\usetikzlibrary{shapes.symbols}
\usetikzlibrary{arrows}
\usetikzlibrary{automata}
\usetikzlibrary{decorations}
\usetikzlibrary{decorations.pathreplacing}
\usetikzlibrary{shapes.symbols}
\usepackage{amsmath,amssymb}
\usepackage{booktabs}
\usepackage[misc]{ifsym}
\usepackage{graphicx}
\usepackage{graphicx}
\usepackage{zref-clever}
\usepackage{hyperref}
\usepackage{color}

\usepackage{csquotes}
\usepackage[shortlabels,inline]{enumitem}
\usepackage{listings,listings-rust}
\usepackage{stmaryrd}

\usepackage{keyrustydl,keybook}

\usepackage%
[disable]
{todonotes}

\newif\ifShowDiff
\newcommand{\diff}[1]{\ifShowDiff\textcolor{blue}{#1}\else#1\fi}

\newif\ifShowApp
\ShowApptrue

\zcRefTypeSetup{section}{cap=true}
\zcRefTypeSetup{appendix}{cap=true,abbrev=true}
\zcRefTypeSetup{figure}{cap=true,abbrev=true}
\zcRefTypeSetup{example}{cap=true}
\zcRefTypeSetup{table}{cap=true}
\zcRefTypeSetup{listing}{cap=true}
\zcRefTypeSetup{definition}{cap=true}

\zcLanguageSetup{english}{
  type=appendix,
  Name-sg-ab = App.,
  name-sg-ab = app.,
  Name-pl-ab = App.,
  name-pl-ab = app.
}
      
\AtBeginDocument{}

\begin{document}
\title{\RustyDL: A Program Logic for \Rust}
%
%
\author{Daniel Drodt(\Letter)\orcidID{0000-0003-3036-8220} \and
  Reiner H\"ahnle\orcidID{0000-0001-8000-7613}
}
\authorrunning{D.\ Drodt and R.\ H\"{a}hnle}
%
\institute{Technische Universit\"at Darmstadt, Darmstadt, Germany\\
\email{\{daniel.drodt,reiner.haehnle\}@tu-darmstadt.de}}
\maketitle
\begin{abstract}
  \Rust is a modern programming language that guarantees memory safety
  and the absence of data races with a strong type system. We present
  \RustyDL, a program logic for \Rust, as a foundation for an
  auto-interactive, deductive verification tool for \Rust. \RustyDL
  reasons about \Rust programs directly on the source code level, in
  contrast to other tools that are all based on translation to an
  intermediate language. A source-level program logic for \Rust is
  crucial for a human-in-the-loop (HIL) style of verification that
  permits proving highly complex functional properties. We discuss
  specific \Rust challenges in designing a program logic and calculus
  for HIL-style verification and propose a solution in each case. We
  provide a proof-of-concept of our ideas in the form of a prototype
  of a \Rust instance of the well-known deductive verification
  tool~\KeY.
  
  \keywords{Deductive verification \and Program logic \and \Rust verification}
\end{abstract}

\section{Introduction}

\Rust is a modern programming language with a strong type system,
guaranteeing memory safety without garbage collection. For this, it
relies on a type system for managing \emph{ownership}, references, and
borrowing, which prevents data races. \Rust became highly popular and
started to being used in the Linux kernel~\cite{LGYWX2024,T2023}. In
such safety-critical use cases, the correctness of programs is
paramount, so formal verification plays an important role.

Over the last years, a variety of tools for formally verifying \Rust
code have been proposed. The general approach of most existing
tools~\cite{ABFGMMPS2022,BL2024a,DJM2022,LHCBSZHPH2023} is to
translate \Rust code and formal specifications into an intermediate
language, such as Viper~\cite{MSS2016} or Why3~\cite{FP2013}, which is
supported by existing verification tools. This intermediate code is
then used to generate first-order verification conditions to be
discharged with SMT solvers.

This translation-based approach permits to implement a deductive
verification tool with relative ease, because it is based on an
existing tool chain. It also offers a high degree of automation. The
downsides of the architecture are: first, one has to trust the (highly
non-trivial) translation from \Rust and its specification to an
intermediate language; second, it is challenging to relate the output
of an SMT solver to the code and specification under verification, in
case a verification attempt fails: a verification engineer cannot
directly interact with the proof or apply proof steps manually.

In the spectrum between full automation of proofs and full control
over the verification process, existing \Rust verification tools are
trending towards the former. At the other end, verification tools such
as \KeY \cite{BBDHLPUW2025} or KIV \cite{KIV20} provide explicit proof
objects and allow their users to inspect and apply any proof step,
while retaining powerful automation capabilities. This is achieved
with a program logic at the \emph{source code} level, typically
axiomatized in a sequent calculus. Such a \enquote{human-in-the-loop}
(HIL), source code-level approach permits fine-grained
control. Consequently, it was possible to verify (and find bugs in)
highly complex, unaltered \Java source code
libraries~\cite{BSUWW2024,DRDBH2015}. This could so far not be
replicated with systems based on translation to an intermediate
language.


Right now, there is no HIL, source code-level deductive verifier for
\Rust. Therefore, a natural research question arises that we answer in
the present paper: Can one transfer HIL, source code verification to
the world of \Rust?  What are the necessary conditions?  How can we
model \Rust's type system and behavior, including ownership and
references, in a source code program logic?  Accordingly, our main
result is the program logic \RustyDL for HIL, source code
verification. We show that a Dynamic Logic with a few familiar
extensions is a natural fit for \Rust verification. As a
proof-of-concept, we also provide a prototype of an implementation
based on the \KeY verifier, called \emph{\Rusty \KeY}.

\zcref[S]{sec:pl} provides background on \Rust and introduces our
program logic for \Rust. In \zcref{sec:calc}, we define the sequent
calculus for \RustyDL. \zcref[S]{sec:related} compares our approach
with related work. We conclude the paper in \zcref{sec:concl}.

\section{The \Rusty Program Logic}%
\label{sec:pl}

\Rust is a complex language with a rich type system and support for
concurrency. It is divided into two sub-languages: \emph{safe \Rust}
limits the developer but provides strong memory safety guarantees,
while \emph{unsafe \Rust} has weaker guarantees but offers more
control. The latter must be marked with the keyword
\lstinline|unsafe|.

We focus on \emph{safe} \Rust because:
\begin{enumerate*}[(1)]
\item most \Rust functions are safe;
\item modeling safe \Rust provides opportunities (and challenges) for
  verification: mechanisms such as separation logic or dynamic frames
  are unnecessary in this setting; and
\item it can be assumed that unsafe \Rust is encapsulated, i.e., it is
  confined to a part of the code that may be analyzed and verified
  separately~\cite{AMPMS2020}.
\end{enumerate*}
Further, we focus here on a subset of \Rust, mainly primitive values,
shared and mutable references, and arrays. Structs and enums can be
modeled similarly to arrays or references. We leave unions, smart
pointers, and traits for future work.

\subsection{The \Rust Programming Language}%
\label{sec:rust}

We briefly review those features of \Rust necessary to follow this
paper. For details and the entirety of \Rust, we refer to
\cite{KN2022,R2024}.
\Rust applications and libraries are organized in units called
\emph{crates}, which can depend on other crates. Each crate is
split into modules containing functions, type definitions, and
implementations. 

The \Rust types important for this paper are \lstinline|bool|, the
integer types \lstinline|i8|,\ldots, \lstinline|i128| for signed
integers, \lstinline|u8|,\ldots, \lstinline|u128| for unsigned
integers, the empty tuple type~\lstinline|()|, and array types such as
\lstinline|[i16; 10]|.  \Rust arrays are of fixed length (here: $10$)
and their length is part of the type.

\Rust's unique feature is the type system's enforcement of
\emph{ownership}, and \emph{references}. Ownership means each value is
\emph{owned} by exactly one \emph{owner} (e.g., a variable or an array
field). The value is removed from memory, or \emph{dropped}, once the
owner goes out of scope. This feature is the basis of \Rust's memory
safety.
Ownership of a value is \emph{transferred} on assignment---one says
the value is \emph{moved}.

\begin{lstlisting}[caption={Shared and mutable references in \Rust},numbers=left,xleftmargin=2em,label={lst:ref},float=t]
let n: u32 = 1;
let x: &u32 = &n;
println!("{}", *x); // prints "1"

fn increase_lower(a: &mut u64, b: &mut u64, amount: u64) {
  let lower = if a < b { a } else { b }; %\label{lst:moving}%
  *lower = *lower + amount;
}
\end{lstlisting}

\begin{example}
  Function \lstinline|increase_lower| in \zcref{lst:ref} takes two
  mutable references \lstinline|a|, \lstinline|b| as arguments,
  compares their values, stores the lower reference in \pv{lower}, and
  increases that value by \pv{amount}. Observe that in \Rust, control
  structures such as conditionals, loops, and blocks are expressions.
  The assignment in \zcref{lst:moving} moves either \pv{a} or \pv{b}
  to \pv{lower}. Hence, one of \pv{a} or \pv{b} is \emph{no longer
    initialized}, while \pv{lower} now contains the moved
  value. Reading from \pv{a} or \pv{b} after \zcref{lst:moving} is a
  compile-time error. This ensures that only initialized memory is
  read.
\end{example}

Types with a small memory footprint, e.g., integer types, do not
follow the move semantics. Their value is \emph{copied} rather than
moved on assignment. In \Rust terminology: They implement the
\lstinline|Copy| trait. The only types in this paper not implementing
this trait are mutable references and arrays of mutable references.

Since accessing a value only through one owner is, in general, too
limiting, \Rust provides references. \emph{Shared (or
  immutable)
  references} (\lstinline|&|) permit read-only access to the
\emph{borrowed} value, while \emph{mutable references}
(\lstinline|&mut|) permit read-and-write access.
As seen in the first three lines of \zcref{lst:ref}, a shared
reference of type \lstinline|u32| is of type \lstinline|&u32|. The
borrowed value \lstinline|1| can be read from reference \pv{x} using
the \emph{dereferencing} operator \lstinline|*x|. While the
borrow~\pv{x} is live, we say \pv{n} is borrowed.
%
%
In case of a mutable reference, the borrowed value cannot merely be
read but also modified using the dereferencing operator, see the
second to last line of the example.

The \Rust compiler---more specifically, its \emph{borrow
  checker}---ensures programs follow certain rules, which imply the
absence of data races:
\begin{enumerate*}[(1)]
\item borrowed owners are not written to directly;
\item mutably borrowed owners are only read through the reference;
\item at any point in the program, each owner is borrowed either
  mutably once or immutably arbitrarily often;
\item borrowed values are not dropped.
\end{enumerate*}

References in \Rust have an associated \emph{lifetime} which is
necessary in complex programs. We do not consider lifetimes in our
logic and omit them here.

In the following, we not only consider entire crates but also
\emph{fragments} of \Rust programs. \zcref[S]{def:frag} defines the
set of legal program fragments, i.e., the \Rust fragments our logic
can deal with.

\begin{definition}[Legal program fragment]\label{def:frag}
  Let $\Crate$ be a valid, i.e., compiler approved, \Rust crate. A
  \emph{legal program fragment} is a sequence of \Rust statements $p$
  in $\Crate$ (and an optional \Rust expression at its end), with
  local variables $\pv{a}_1,\ldots,\pv{a}_n$ of \Rust types
  $\rust{T}_1,\ldots,\rust{T}_n$ such that inserting the function
  \begin{center}
    \lstinline|fn f(mut a$_1$: T$_1$, $\ldots$, mut a$_n$: T$_n$) { { $p$ }; }|
  \end{center}
  into the entry point of $\Crate$ yields a legal program according to
  the \Rust compiler.
\end{definition}

The fragment $p$ is wrapped in a block expression. We add the
semicolon to ensure that the function returns nothing, simplifying
type checking.

\subsection{Rusty Dynamic Logic}\label{sec:dl}

\emph{Dynamic logic} is a program logic suggested in~\cite{P1976} and
being highly suitable for program verification~\cite{Reif65}. We
present a version of dynamic logic tailored to \Rust called \RustyDL,
which is an extension of typed first-order logic. Our definitions are
based on those for \JavaDL, a dynamic logic for \Java, as presented
in~\cite{ABBHSU2016}.

\begin{figure}[t]
  \centering
  \begin{tikzpicture}[arrows=->]
    \node (any) at (5, 2.5) {$\Any$};
    \node (bool) at (0, 1.5) {$\bool$};
    \node (int) at (1.5, 1.5) {$\Int$};
    \node (unit) at (3, 1.5) {$\unit$};
    \node (arrays) at (5, 1.5) {array sorts};
    \node (field) at (7, 1.5) {$\Field$};
    \node (place) at (8.2, 1.5) {$\Place$};
    \node (refs) at (10, 1.5) {reference sorts};
    \node (never) at (5, 0.5) {$\neverTy$};

    \draw (never)--(bool);
    \draw (never)--(int);
    \draw (never)--(unit);
    \draw (never)--(arrays);
    \draw (never)--(field);
    \draw (never)--(refs);
    \draw (never)--(place);
    \draw (bool)--(any);
    \draw (int)--(any);
    \draw (unit)--(any);
    \draw (arrays)--(any);
    \draw (field)--(any);
    \draw (place)--(any);
    \draw (refs)--(any);
  \end{tikzpicture}
  \caption{The sort hierarchy of \RustyDL}
  \label{fig:sorts}
\end{figure}
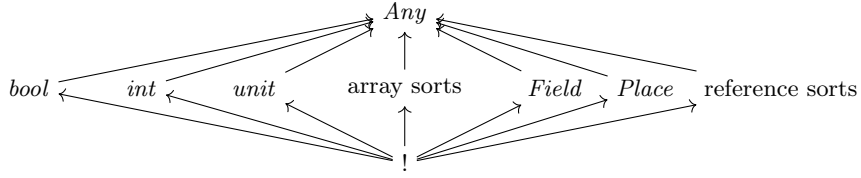

\zcref[S]{fig:sorts} shows \RustyDL's sort hierarchy, i.e., the
hierarchy of logical types. $\Any$ is the super-sort of all sorts,
while the \enquote{never sort} $\neverTy$ is the bottom sort no term
can have. The type \lstinline|bool| is mapped to $\bool$, the unit
type \lstinline|()| to $\unit$. Signed and unsigned integer types are
mapped to mathematical integers $\Int$ (see \zcref{sec:int} for how we
achieve an accurate model).
For each array type \lstinline|[T;N]| we create an array sort
$\arrTy{T}{N}$.
Further, for each sort $A$, we define a shared reference sort
$\RefS{A}$ and a mutable reference sort $\RefM{A}$. \Rust types
\lstinline|&T| and \lstinline|&mut T| map to the reference sorts in
the expected way.

We demand of the signature that it contains the $0$-ary predicates
$\keytrue$ and $\keyfalse$, the binary predicate $\keyeq(\Any,\Any)$
for equality, all usual integer functions and predicates such as
$+, -, \cdot, >$, and the constants $\keybooltrue, \keyboolfalse$ of
sort $\bool$. \diff{We use both $\keytrue$ and $\keybooltrue$ to
  differentiate between the tautology and the true boolean value, as
  the former is a formula and the latter a term; hence, they may
  appear in different scenarios. Analogous for $\keyfalse$ and
  $\keyboolfalse$.}

\RustyDL extends the set of classical first-order logic terms with
\emph{program variables}, e.g., $\pv{a}: \Int$, where $\Int$ is
\pv{a}'s sort. These are similar to constants, but their value is
state-dependent. Unlike first-order logical variables, program
variables may appear in programs, but one cannot quantify over program
variables.

We extend the set of first-order formulas with the modal operators
$\dia{p}\phi$ (\enquote{diamond}) and $\dlbox{p}\phi$ (\enquote{box}),
for program fragment $p$ and formula $\phi$. Intuitively, for
deterministic programs, $\dia{p}\phi$ expresses that $p$ terminates,
and, in the state resulting from this execution, $\phi$ holds,
i.e.~total correctness. Conversely, $\dlbox{p}\phi$ holds iff $p$ does
not terminate, or $\phi$ holds in the resulting state, i.e.~partial
correctness.

A \emph{Kripke structure} $\Kstruct=(\domain,\interp,\States,\rho)$
over domain~$\domain$ and interpretation~$\interp$ formalizes the
state changes expressed by the modality formulas, ordering a set of
states~$\States$ according to possible state transitions~$\rho$
through programs.  States in~$\States$ map program variables to
elements in~$\domain$. $\interp$, as in FOL, evaluates functions and
relations. For a program fragment $p$, $\rho(p)$ is a partial function
from $\States$ to $\States$ that defines the state change caused by
$p$. This change is only defined if $p$ terminates normally, \diff{see
  \zcref{def:kripke}}.
\diff{\begin{definition}[\RustyDL Kripke structure]\label{def:kripke}
  Let $\Crate$ be a valid \Rust crate. A \emph{\RustyDL Kripke
    structure} for a signature derived from $\Crate$ is a pair
  \[
    \Kstruct=(\States,\rho)
  \]
  where
  \begin{itemize}
  \item $\States$ is an infinite set of first-order structures
    $s=(\Domain,\interp)$, called \emph{states}, closed under the
    following property: all states coincide in their domain $\Domain$
    and interpretation $\interp$ of function and predicate symbols.
  \item $\rho$ is a function mapping a legal program fragment $p$ to a
    partial function $\States\rightharpoonup\States$ with
    $\rho(p)(s_1)=s_2$ iff evaluating $p$ in state $s_1$ terminates
    normally in $s_2$. If $p$ does not terminate normally starting in
    $s_1$, $\rho(p)(s_1)$ is undefined. Note that we assume \Rust
    programs to be deterministic.
  \end{itemize}
\end{definition}}

For brevity, the transition function~$\rho$ is left undefined, as it
depends on a formal semantics for \Rust.
Our calculus rules provide an axiomatic semantics of~$\rho$.
\zcref[S]{def:sem} gives the semantics of terms and formulas,
including program variables and modalities, relative to a Kripke
structure $\Kstruct$.

\begin{definition}[Semantics of \RustyDL terms and formulas]\label{def:sem}
  Let $\Crate$ be a valid \Rust crate, $\VSym$ the set of logic
  variable symbols, $\Kstruct=(\domain,\interp,\States,\rho)$ a Kripke
  structure modeling \Rust programs, $s\in\States$ a state, and
  $\beta:\VSym\rightarrow\domain$ a variable
  assignment.

  For every \RustyDL term $t$ we define its evaluation inductively by
  the evaluation function $\valS$ exactly as in first-order logic,
  with evaluation of program variables defined as
  $\valS(\pv{a})=s(\pv{a})$ for each program variable~\pv{a}.
  
  For every \RustyDL formula $\phi$ we define satisfaction, denoted
  $\valid\phi$, exactly as in first-order logic with two new cases:
  \begin{enumerate}
  \item $\valid\dlbox{p}\phi$ iff $\rho(p)(s)$ is undefined or
    $\rho(p)(s)=s'$ and $\validSPrime\phi$,
  \item $\valid\dia{p}\phi$ iff $\rho(p)(s)=s'$ exists and
    $\validSPrime\phi$.
  \end{enumerate}
\end{definition}

In \RustyDL one can express a large variety of properties of \Rust
code.  Partial correctness of program $p$ w.r.t.\ pre- and
post-conditions $\pre,\post$ is written as
$\pre\rightarrow\dlbox{p}\post$. This formula is equivalent to the
Hoare triple $\{\pre\}\,p\,\{\post\}$~\cite{H1969} and it is valid iff
for every state $s$ such that $\pre$ is satisfied in~$s$, evaluating
program $p$ in $s$ either does not terminate or in the resulting
state~$s'$, $\post$ is satisfied. In contrast,
$\pre\rightarrow\dia{p}\post$ expresses total correctness.
%
Unlike Hoare logic, dynamic logic is closed under all
operations. Hence, e.g., $\post$, may contain further modalities
permitting to express program equivalence, refinement,
etc.~\cite{Reif65}.

To obtain a usable \RustyDL calculus that can model state changes
efficiently, we provide a more fine-grained state transition mechanism
than the modalities called \emph{update}.  \zcref[S]{def:upd} extends
\JavaDL \emph{updates} \cite{ABBHSU2016} with the \emph{mutating
  updates}.

\begin{definition}[Update, Update Application]\label{def:upd}
  We define the set $\Updates$ inductively by (the second clause is
  new compared to \cite{ABBHSU2016}):
  \begin{enumerate}
  \item $(\pv{a}\upd t)\in\Updates$ for each program variable
    $a:A$ and each term $t$ of sort $A$,
  \item $(t_1\dupd t_2)\in\Updates$ for each term
    $t_1$ of sort $\RefM{A}$ and $t_2$ of sort $A$,
  \item $\skipUp\in\Updates$,
  \item $(u_1\parupd u_2)\in\Updates$ for all updates
    $u_1,u_2\in\Updates$,
  \item $(\applyUp{u_1}{u_2})\in\Updates$ for all updates
    $u_1,u_2\in\Updates$.
  \end{enumerate}
  We extend the inductive definition of terms with term
  $\applyUp{u}{t}$ of sort $A$ for all updates $u\in\Updates$ and
  terms $t$ of sort $A$. The inductive definition of formulas is
  extended with formula $\applyUp{u}{\phi}$ for all updates
  $u\in\Updates$ and formulas $\phi$.
\end{definition}

Updates are \emph{applied} syntactically to terms and formulas. Update
application changes the state wherein a term (formula) is
evaluated. Updates allow us to realize the decomposition of a large
program fragment into a series of small, deterministic state changes
(see \zcref{def:upd-sem} below).

\begin{example}
  Let $\pv{x},\pv{y}$ be program variables of type \lstinline{i32} and
  \pv{z} of type \lstinline|&mut i32|. We give examples of updates
  with their intuitive meaning:
  \begin{itemize}
  \item $\skipUp$ is an update expressing no state change.
  \item An elementary update $(\pv{x}\upd\pv{x}+1)$ expresses a state
    change where program variable $\pv{x}$ is incremented by one. The
    update application $\applyUp{\pv{x}\upd\pv{x}+1}{\pv{x}}$ is
    semantically equivalent to the expression
    $\pv{x}+1$.
  \item The \emph{parallel} update
    $(\pv{x}\upd\pv{y}\parupd\pv{y}\upd\pv{x})$ expresses a state
    change where $\pv{x}$ and $\pv{y}$ swap values. Such parallel
    updates express state changes with no interference among
    them. 
    Multiple updates to the same variable can occur in one parallel
    update; then, the last one overwrites the earlier ones.
  \item $(\pv{z}\dupd \pv{y}-1)$ is a \emph{mutating} update,
    expressing that the location borrowed by~$\pv{z}$ is set to
    $\pv{y}-1$.
  \end{itemize}
\end{example}

Mutating updates are crucial for representing mutable references in
our logic. To model the latter in \RustyDL, we need to involve not
merely values, but also the borrowed location (the \emph{lender}). To
this end, we use the sort $\Place$ (the notion is taken from the \Rust
compiler~\cite{R2025}). For each program variable~\pv{x}, we define a
unique constant $\place{\pv{x}}:\Place$. To model the location of an
array field, we add a function
$\arrplaceSym:\RefM{\arrTy{A}{n}}\times\Int\rightarrow\Place$ for each
sort~$A$ and $n\in\mathbb{N}$, intuitively, mapping each field to its
place. We write $\arrplace{t}{t'}$ for $\arrplaceE{t}{t'}$. Thus, the
term $\refM{pl}$ is of sort $\RefM{A}$, where $pl:\Place$ is the place
of the lender of sort~$A$. How these mutable references are used is
shown in \zcref{sec:refs}.

\begin{example}
  Let $\pv{x}:\Int$ and $\pv{a}:\arrTy{\Int}{4}$ be program
  variables. The mutable reference $\refM{\place{\pv{x}}}$ expresses
  that \pv{x} is mutably borrowed and
  $\refM{\arrplace{\refM{\place{\pv{a}}}}{1}}$ expresses that the
  field at index $1$ of array~\pv{a} is mutably
  borrowed. 
\end{example}

Mutating updates are the central extension of our logic. They fit
naturally into our dynamic logic and allow for handling mutable
references with little conceptual overhead. Alternatives, such as
explicitly modeling memory or the set of active loans, rapidly become
unwieldy and computationally complex when used in a logic.

\zcref[S]{def:upd-sem} formalizes the semantics of updates and their
applications that was described above intuitively.

\begin{definition}[Semantics of \RustyDL updates]\label{def:upd-sem}
  Let $\Crate$ be a valid \Rust crate, $\Kstruct=(\States,\rho)$ a
  Kripke structure for $\Crate$, $s\in\States$ a state, and
  $\beta:\VSym\rightarrow D$ a variable assignment. The valuation
  function $\valS:\Updates\rightarrow(\States\rightarrow\States)$ is
  defined as:
  \begin{align*}
    \valS(\pv{a}\upd t)(s')(\pv{b})=&
                                      \begin{cases}
                                        \valS(t) &\text{if}~\pv{b}=\pv{a}\\
                                        s'(\pv{b})&\text{otherwise}
                                      \end{cases}\;\text{for all }s'\in\States,\pv{b}\in\PVSym\\
    \valS(t_1\dupd t_2)(s')(\pv{b})=&
                                      \begin{cases}
                                        \valS(t_2)&\text{if}~\valS(t_1)=\valS(\refM{\place{\pv{b}}})\\
                                        s'(\pv{b})&\text{otherwise}
                                      \end{cases}\\
                                    &\text{for all}~s'\in\States,\pv{b}\in\PVSym\\
    \valS(\skipUp)(s')=&\,s' \quad \text{for all }s'\in\States\\
    \valS(u_1\parupd u_2)(s')=&\,\valS(u_2)(\valS(u_1)(s')) \quad \text{for all } s'\in\States\\
    \valS(\applyUp{u_1}u_2)=&\,\valSPrime(u_2) \text{ where }s'=\valS(u_1)(s)
  \end{align*}
  We extend the inductive definition of the semantics of \RustyDL
  terms and formulas (\zcref{def:sem}) with
  \[
    \valS(\applyUp{u}{t})=\valSPrime(t)\text{ where } s'=\valS(u)(s)
  \]
  and
  \[
    \valid\applyUp{u}{\phi} \text{ iff } \validSPrime\phi \text{ where
    }s'=\valS(u)(s)
  \enspace.\]
\end{definition}

\begin{example}
  Let $\pv{x}$ be a program variable of type $\rust{i32}$. Clearly,
  $\pv{x}\keyeq 0\rightarrow\pv{x}>0$ is unsatisfiable. However,
  applying the update $\pv{x}\upd\pv{x}+1$ to the right-hand side
  makes the formula valid:
  \[
    \models\pv{x}\keyeq 0\rightarrow\applyUp{\pv{x}\upd\pv{x}+1}{(\pv{x}>0)}.
  \]
  Compare the update above with the following modality:
  \[
    \pv{x}\keyeq 0\rightarrow\dlf{x = x + 1;}{(\pv{x} > 0)}.
  \]
  
  It is instructive to observe the difference: Updates are
  \emph{total} state changes, they are guaranteed to terminate
  normally. Hence, they are similar to explicit
  substitutions. Modalities, on the other hand, may not
  terminate. Addition may lead to an overflow, causing exceptional
  termination: $\rho(p)$ is partial.
\end{example}

\section{\RustyDL Calculus}\label{sec:calc}

Recall that functional total correctness of a program fragment $p$,
relative to a precondition $\pre$ and postcondition $\post$, is
expressed by validity of the \RustyDL formula
$\pre\rightarrow\dia{p}\post$. This can be proven in a sequent
calculus for \RustyDL: Validity is shown by deriving the sequent
\enquote{$\seq{}{\pre\rightarrow\dia{p}\post}$}.

The calculus given below is sound: All rules in this section have been
proven sound relative to a formal semantics~\cite{LAGC23} in
\diff{ongoing work}.

\subsection{Challenges in the Design of a Sequent Calculus for
  \RustyDL}%
\label{sec:challenges}

We model \Rust's unique features and its semantics in terms of rules
of a program logic, striving to retain a human-readable format to
enable human interaction. For this, we work on the only slightly
simplified and normalized high-level intermediate representation (HIR)
used in the \Rust compiler~\cite{R2025}. This representation is close
to source-level \Rust but with macros expanded and some structures
like the different forms of loop normalized.

We identify five core features of \Rust our logic must address:
\begin{enumerate*}[(1)]
\item ownership,
\item integer semantics,
\item data types, e.g., arrays, structs,
\item shared and mutable references, including references to array
  fields, and
\item the potentially unbounded behavior of function calls and loops.
\end{enumerate*}
For space reasons, we only present an essential subset of the \Rust
fragment supported by our logic and implementation. The subset
discussed in the following demonstrates how our logic and calculus
models \Rust's features while being the necessary minimum for complex
problems.

\subsection{Symbolic Execution Calculus}

The sequent calculus contains the usual propositional and first-order
rules, as well as a relatively complete rule set for mathematical
integers. Since formulas may contain modalities, we require a way to
simplify and decompose the program fragments contained in them until
only first-order formulas remain. We follow
\cite{ABBHSU2016,BBDHLPUW2025} in the design of rule schemata that
implement \emph{symbolic execution}.

Consider an \rust{if}-\rust{else} expression. Its execution branches
into two paths, depending on the value of its guard. Because we work
with \emph{symbolic} program variables, the calculus must cover all
possible paths and split the current proof goal.

As seen in rule~$\ruleName{ifElseSplit}$ below, the rules contain
schematic variables: $\Gamma$, $\Delta$ stand for sets of
\RustyDL-formulas, $\phi$ for one \RustyDL-formula, $\updateContext$
for an optional
update, $\mathit{se}$ for a \enquote{simple expression} (a literal or
program variable), \rust{\{$\mathit{be}_0$\}} and
\rust{\{$\mathit{be}_1$\}} for block expressions. Additionally, the
program modalities contain a \emph{context} of the program fragment in
focus, with \emph{inactive program prefix} $\pi$ and \emph{remaining
  program} $\omega$. Typically, $\pi$ matches opening braces and
$\omega$ the remaining statements and closing braces. This context
allows the rule below to match any program fragment where the
expression to be executed next is an \rust{if}-\rust{else}, even if
further statements follow or it is the start of a block expression.
All schematic variables are instantiated by concrete objects of the
corresponding kind upon rule application.
\[
  \seqRule{ifElseSplit}%
  {\sequent{\updateContext(\mathit{se}\keyeq\keybooltrue)}{\updateContext\ptstub{\{$\mathit{be}_0$\}}} \\
    \sequent{\updateContext(\mathit{se}\keyeq\keyboolfalse)}{\updateContext\ptstub{\{$\mathit{be}_1$\}}}}
  {\sequent{}{\updateContext\ptstub{if
        $\mathit{se}$ \{$\mathit{be}_0$\} else \{$\mathit{be}_1$\}}}}
\]

The rule models symbolic execution of a program starting with an
\rust{if}-\rust{else} expression that has a simple (hence, side-effect
free) guard. The proof goal in the conclusion is reduced to the two
premises corresponding to each execution path. The sequents in the
premises express that we know the value of the guard $\mathit{se}$ in
each branch: In the then-branch 
we know $\mathit{se}$ to be $\keybooltrue$, in the else-branch
to be $\keyboolfalse$.
%
Matching on expressions (in contrast to statements) allows us to
handle assignments occurring in statements (\lstinline|x = y;|) and,
for example, as a block's value (\lstinline|{x = y}|), with a single
rule.

\diff{If the \rust{if}'s condition is a complex expression, i.e.,
  neither a literal nor a program variable, we symbolically execute
  this condition by introducing a fresh auxiliary variable. In the
  rule $\ruleName{unfoldIfElse}$, $\mathit{nse}$ stands for a
  \enquote{non-simple expression} and \pv{x} is a so far unused
  program variable.
  \[
  \seqRule{unfoldIfElse}%
  {\sequent{}{\updateContext\ptstub{\{let x = $\mathit{nse}$; if x \{$\mathit{be}_0$\} else \{$\mathit{be}_1$\}\}}}}
  {\sequent{}{\updateContext\ptstub{if
        $\mathit{nse}$ \{$\mathit{be}_0$\} else \{$\mathit{be}_1$\}}}}
\]
}

To model state changes in the calculus, we employ updates. The rule
$\ruleName{copy}$ below matches on any assignment of a simple
expression to a program variable. Applying the rule adds an update in
front of the modality that represents the assignment's state change
and replaces the original expression of the assignment with the unit
expression \lstinline|()|.
Unsurprisingly, the following rule only works for literals or
variables of \lstinline|Copy| type, i.e., without move
semantics. Assignments causing a move need a separate rule.
%
\[
  \seqRule{copy}%
  {\sequent{}{\updateContext\upl\pv{x}\upd\mathit{se}\upr\ptstub{()}}}
  {\sequent{}{\updateContext\ptstub{x = $\mathit{se}$}}}\quad
  \mbox{$\mathit{se}$ is a literal or implements \lstinline|Copy|}
\]

Subsequent applications of the $\ruleName{copy}$ rule accumulate
updates in front of the modality that represent the state changes on
one specific path through the control flow of a program fragment. The
accumulated updates are simplified and ultimately applied to the
postcondition. In consequence, the calculus contains \emph{update
  simplification} rules. For instance, applying an elementary update
$\pv{x}\upd t$, with a program variable $\pv{x}$ and a term $t$ to the
program variable $\pv{x}$, corresponds to substituting $\pv{x}$ with
$t$. In our notation:
\[
  \ruleName{applyOnPV} \quad (\applyUp{\pv{x}\upd t}{\pv{x}}) \leadsto t
\]

See \diff{\ifShowApp\zcref{app:upd-rules}\else\cite{DH2026TR}\fi} for
the complete set of update simplification rules.

To get rid of the unit expressions left by rule~$\ruleName{copy}$ and
expressions like \rust{2} in \rust{\{2; a + b\}}, we use
rule~$\ruleName{simpleExprStmt}$. This rule removes expression
statements consisting of a simple expression, as these cannot affect the
program state.
\[
  \seqRule{simpleExprStmt}%
  {\sequent{}{\updateContext\ptstub{}}}
  {\sequent{}{\updateContext\ptstub{$\mathit{se}$;}}}
\]

\begin{example}
  Let $\pv{x}$, $\pv{y}$, $\pv{z}$ be program variables of type
  \rust{i32}.  Consider the sequent
  $\seq{\pv{x}\keyeq\pv{z}}{\dlf{x = x + y; y = x -
      y;}{(\pv{y}\keyeq\pv{z})}}$. Applying $\ruleName{copy}$ and then
  $\ruleName{simpleExprStmt}$ twice
  yields the sequent
  $\seq{\pv{x}\keyeq\pv{z}}{\applyUp{\pv{x}\upd\pv{x}+\pv{y}}{\applyUp{\pv{y}\upd\pv{x}-\pv{y}}{\dlf{}{(\pv{y}\keyeq\pv{z})}}}}$
  (ignoring overflow, to be discussed in \zcref{sec:int}). It can be
  simplified (rule not shown) to
  $\seq{\pv{x}\keyeq\pv{z}}{\applyUp{\pv{x}\upd\pv{x}+\pv{y}}{\applyUp{\pv{y}\upd\pv{x}-\pv{y}}{(\pv{y}\keyeq\pv{z})}}}$,
  after update application and simplification it becomes
  $\seq{\pv{x}\keyeq\pv{z}}{(\pv{x}+\pv{y}-\pv{y}\keyeq\pv{z})}$, a
  valid first-order sequent.
\end{example}

\subsection{Ownership}%
\label{sec:own}

Consider an assignment \lstinline|x = y|, where \pv{y} is not of a
type implementing \lstinline|Copy|, for example, a mutable
reference.
Then the value of \pv{y} is moved out of \pv{y} to \pv{x}. We model
this behavior in the logic: After the move, \pv{x}'s value should not
be accessible via \pv{y} in a formula.
%
One way to model this in logic is with explicit \emph{uninitialized
  values}. The downside is that then every variable on each access
must be tested whether it contains such a special value. This is
similar to the problems arising when using many-valued logic to model
undefined values~\cite{H2005}. It is more efficient to rely instead on
\emph{underspecification} and simply use Skolemization (called
anonymizing update in \cite{ABBHSU2016}). The rule $\ruleName{move}$
is similar to $\ruleName{copy}$, but it updates \pv{y} to a fresh
constant $c$ of matching type: While~\pv{y} still has \emph{some}
value, we know nothing about it besides its sort.
%
\[
  \seqRuleC{move}%
  {\sequent{}{\updateContext\upl\pv{x}\upd\pv{y}\parupd\pv{y}\upd
      c\upr\ptstub{()}}} {\sequent{}{\updateContext\ptstub{x = y}}}
  {$\pv{y}: A$, $A$ is not \lstinline|Copy| and $c:
    A$ a fresh constant}
\]

\subsection{Handling Integers and Overflow}\label{sec:int}

\Rust provides multiple compilation modes with differing levels of
optimization, the most prominent being called \emph{debug} and
\emph{release}. The former contains more checks at the cost of
performance. These modes also affect integer operations.

An expression like $\rust{a + b}$ may overflow which in \Rust's
semantics is unintended and causes a panic in debug mode~\cite{R2024}.
In release mode, overflow checks are omitted and the value wraps
around. For verification, we focus on the debug semantics, so \RustyDL
must show safety: the absence of overflows.
Rule~$\ruleName{assignAddU32}$ matches on assigning the sum of two
simple expressions to program variable \pv{x} with \Rust type
\lstinline|u32|. The rule splits a proof into two cases. If the
result of the addition is in range of the \lstinline|u32| type, the
sum is translated to an update. Otherwise, a panic is raised.
\[
  \seqRuleW{assignAddU32}%
  {\sequent{\updateContext\,\dlfunc{inU32}(\mathit{se}_1+\mathit{se}_2)}{\updateContext\upl\pv{x}\upd\mathit{se}_1+\mathit{se}_2\upr\ptstub{()}}\\
    {\sequent{\updateContext\neg\dlfunc{inU32}(\mathit{se}_1+\mathit{se}_2)}{\updateContext\ptstub{panic!()}}}}
  {\sequent{}{\updateContext\ptstub{x = $\mathit{se}_1$+$\mathit{se}_2$}}}
\]

The semantics of the $\dlfunc{inU32}$ predicate is a bounds check
between the constants \rust{u32::MIN} ($=0$) and \rust{u32::MAX}
($=2^{32}-1$).

\subsection{Rules for References}\label{sec:refs}

Recall that a program fragment $p$ is only valid if the \Rust
compiler---including the borrow checker---accepts it. Thus, our
calculus needs not to ensure that $p$ adheres to the borrow checker's
rules, instead, it can rely on its guarantees.

How to model references in \RustyDL? One approach is to add a reserved
program variable, or a ghost variable, call it \pv{loans}, that stores
which variables borrow from which others and whether these borrows are
shared or mutable. Thus, \pv{loans} represents the set of active loans
at any moment. The reference rules may then be explicitly introduced,
if necessary, as axioms or rules over \pv{loans}.
However, an explicit model of active loans must be kept up-to-date
after \emph{every state change}. If we encounter a statement as simple
as \rust{x = y}, we cannot use $\ruleName{copy}$, because we also
might have to update \pv{loans}: What if \pv{x} or \pv{y} are
borrowed? Then all active loans for these variables must be removed
from \pv{loans}.
This approach does not benefit from \Rust's borrow checker, but in
effect re-implements it.

Therefore, we do not explicitly model the active loans, but express
references as logical terms.
For shared references, this is simple. On encountering
\lstinline{x = &y}, we know that the value \pv{x} is borrowing cannot
be changed while \pv{x} is live. Hence, one disregards where the value
is coming from and simply works with \pv{y}'s value:
\[
  \seqRule{borrowShared}%
  {\sequent{}{\updateContext\upl\rust{x}\upd\refS{\rust{y}}\upr\ptstub{()}}}%
  {\sequent{}{\updateContext\ptstub{x = \&y}}}
\]

The current value of \pv{y} is held by $\Gamma,\Delta$, or
$\updateContext$. Any further change to \pv{y} (which must necessarily
come after \pv{x} is dropped or re-assigned) does not affect
\pv{x}. Since \pv{x} is not equivalent to \pv{y}, we do not assign
\pv{y} to \pv{x} but $\refS{\pv{y}}$, a shared reference
term. Function $\refSSym$ takes a term of type $A$ and returns a term
of type $\RefS{A}$.

\begin{example}
  Consider the sequent
  $\seq{}{\dlf{y = 2; x = \&y}{(\pv{x}\keyeq\refS{2})}}$. Applying
  first rule $\ruleName{copy}$, then $\ruleName{borrowShared}$ results
  in
  \[
    \seq{}{\applyUp{\pv{y}\upd 2}{\applyUp{\pv{x}\upd\refS{\pv{y}}}{\dlf{}{(\pv{x}\keyeq\refS{2})}}}}
  \enspace.\]
\end{example}

Reading values of shared references is done by function
$\derefS_A:\RefS{A}\rightarrow A$ for each sort $A$ (omitted in the
following for readability).  Rule $\ruleName{derefShared}$ takes an
assignment of the dereference of \pv{y} to \pv{x} and translates it to
an update with~$\derefS$. The simplification rule
$\ruleName{derefOfSharedRef}$ then translates the dereference of a
shared reference of term~$t$ back to $t$.

\noindent\begin{minipage}[c][][t]{.5\linewidth}
  \[
    \seqRuleC{derefShared}
    {\sequent{}{\updateContext\upl\pv{x}\upd\derefS(\pv{y})\upr\ptstub{()}}}
    {\sequent{}{\updateContext\ptstub{x = *y}}}
    {$\pv{x}:A,\pv{y}:\RefS{A}$, $A$ is \lstinline|Copy|}
  \]
\end{minipage}\hfill
\begin{minipage}[c][][t]{.4\linewidth}
  \[
    \rwRule{derefOfSharedRef}{\derefS(\refS{t})}{t}
  \]
\end{minipage}
\vspace{1ex}

As shared references only consider the value, their treatment is
exactly the same for shared references to arrays.

Mutable references differ in that we cannot treat them as the borrowed
value at time of the borrow: If \rust{x} is a mutable borrow of
\rust{y}, there might be an assignment such as \rust{*x =
  $\mathit{se}$}, which changes not \rust{x}'s value but
\rust{y}'s. Thus, \rust{x}'s value contains the lender and we treat
assignments to \rust{*x} as assignments to this place.

Consider borrowing variables. When encountering a mutable borrow, for
example, in \rust{x = \&mut y}, applying rule~$\ruleName{borrowMut}$
adds an update for $\pv{x}$ with $\refM{\place{\pv{y}}}$, representing
that \pv{x} borrows \pv{y}: 
\[
  \seqRule{borrowMut}%
  {\sequent{}{\updateContext\upl\rust{x}\upd\refM{\place{\rust{y}}}\upr\ptstub{()}}}%
  {\sequent{}{\updateContext\ptstub{x = \&mut y}}}
\]

To read the value borrowed by a mutable reference, we add the unary
state-dependent operator $\derefM_A$, which takes a term of sort
$\RefM{A}$ and is itself of sort~$A$ and declare a corresponding
lookup rule \( \ruleName{derefOfMutReference}\)
\( \derefM(\refM{\place{\pv{x}}})\leadsto \pv{x}.  \) The operator is
state-dependent to reflect any update expressing a state change. This
might be necessary when \pv{x} was mutated. To update a
variable~\pv{x} appearing in $\derefM$ but keep other state changes
possibly affecting the lender, we use the rule
\(\ruleName{applyElemOnDeref}\; \applyUp{u_1\parupd\pv{x}\upd t\parupd
  u_2}{\derefM(\pv{x})} \leadsto \applyUp{u_1\parupd
  u_2}{\derefM(t)}.\) 

Additionally, we require a rule for mutating a value through the
mutable reference~$\pv{x}$. We want to update not the value of
$\pv{x}$, but the \emph{place of the lender}~$\place{\pv{y}}$. This is
why we introduced \emph{mutating updates} $t_1\dupd t_2$, expressing
that term $t_2$ is written to the place borrowed by $t_1$.
%
\[
  \seqRule{mutate}%
  {\sequent{}{\updateContext\upl\rust{x}\dupd\mathit{se}\upr\ptstub{()}}}%
  {\sequent{}{\updateContext\ptstub{*x = $\mathit{se}$}}}
\]

\begin{example}\label{exmpl:mutate}
  Consider the sequent
  $\seq{}{\dlf{x = 1; y = \&mut x; *y = 3;}{\phi}}$ with program
  variables $\pv{x}: \Int$, $\pv{y}:\RefM{\Int}$. Applying rules
  $\ruleName{copy}$, $\ruleName{borrowMut}$, and $\ruleName{mutate}$
  results in
  \[
    \seq{}{\applyUp{\pv{x}\upd
        1}{\applyUp{\pv{y}\upd\refM{\place{\pv{x}}}}{\applyUp{\pv{y}\dupd
            3}{\dlf{}{\phi}}}}}\enspace.
  \]
\end{example}

To simplify such mutating updates, we declare two simplification
rules. The first, $\ruleName{applyOnMutating}$, applies an update $u$
to a mutating update. Unlike for elementary updates \diff{\ifShowApp
    (cf.\ \zcref{tab:upd-simp} in \zcref{app:upd-rules})\else
    (see~\cite{DH2026TR} for details)\fi}, $u$ is applied to both
sides of~$\dupd$. Rule $\ruleName{mutatingToElementary}$ transforms a
mutating update, where the left-hand-side is known to be
$\refM{\pv{a}}$ for some program variable $\pv{a}$, to an elementary
update $\pv{a}\upd t$.
\vspace{-1ex}

\noindent\begin{minipage}[b]{.45\linewidth}
  \[
    \rwRule{applyOnMutating}{\upl u\upr(t_1\dupd t_2)}{\upl u\upr
      t_1\dupd\upl u\upr t_2}
  \]
\end{minipage}\hfill
\begin{minipage}[b]{.45\linewidth}
  \[
    \rwRule{mutatingToElementary}{\refM{\place{\rust{a}}}\dupd t}{\pv{a}\upd t}
  \]
\end{minipage}

\begin{example}
  Continuing \zcref{exmpl:mutate}, simplifying the first two updates
  leads to
  \[
    \seq{}{\applyUp{\pv{x}\upd 1\parupd\pv{y}\upd\refM{\place{\pv{x}}}}{\applyUp{\pv{y}\dupd 3}{\dlf{}{\phi}}}}
  \]
  Because the place $\place{\pv{x}}$ is a \emph{constant}, it is
  unaffected by updates. Further simplification yields
  \( \seq{}{\applyUp{\pv{x}\upd
      1\parupd\pv{y}\upd\refM{\place{\pv{x}}}\parupd\refM{\place{\pv{x}}}\dupd
      3}{\dlf{}{\phi}}}.  \) Using $\ruleName{mutatingToElementary}$,
  we get
  \( \seq{}{\applyUp{\pv{x}\upd
      1\parupd\pv{y}\upd\refM{\place{\pv{x}}}\parupd\pv{x}\upd
      3}{\dlf{}{\phi}}} \), where we see that $\pv{y}\dupd 3$ is
  replaced by $\pv{x}\upd 3$. We finally simplify away the overridden
  update to
  \(\seq{}{\applyUp{\pv{y}\upd\refM{\place{\pv{x}}}\parupd\pv{x}\upd
      3}{\dlf{}{\phi}}}\).
\end{example}

\subsection{Modeling Array Access}%
\label{sec:array}

To model array access we use two functions for each array
type~$\arrTy{S}{n}$: Function
$\arrget{S}{n}:\arrTy{S}{n}\times\Field\rightarrow S$ takes an array
and a term of type $\Field$ and evaluates it to the value of the array
at the index represented by the field. Function
$\arrset{S}{n}:\arrTy{S}{n}\times\Field\times S
\rightarrow\arrTy{S}{n}$ takes an array, a field for index $i$, and a
value $t$ and creates a new array term with the value at the index $i$
overwritten with $t$.
The sort $\Field$ allows abstraction over indices: Function
$\arridx:\Int\rightarrow\Field$ turns an integer into a field. Sort
$\Field$ is also used for tuple and struct fields.
The present approach follows \JavaDL and the theory of
arrays~\cite{M1962}. The difference is that in \RustyDL the only
possible aliasing is over the array index, not the array
entries. Array access in programs is symbolically executed: If the
index is out of bounds, we continue with a panic; otherwise, we model
the successful access with an update.

\diff{\begin{example}\label{example:arr}
  Let $\pv{a}:\arrTy{\Int}{4}$ and $\pv{i},\pv{j}: \Int$ be program
  variables. The term
  \[
    \arrget{\Int}{4}(\arrset{\Int}{4}(\arrset{\Int}{4}(\pv{a},\arridx(\pv{i}),
    1), \arridx(\pv{j}), 2),\arridx(\pv{i}))
  \]
  represents an access of \pv{a} at index \pv{i} after first setting
  the value at index \pv{i} to $1$, then writing $2$ at index
  \pv{j}. Assuming \pv{i} and \pv{j} are valid indices, the value of
  this term is $2$ iff $\pv{i}\keyeq\pv{j}$, and $1$ otherwise.
\end{example}

We pretty-print a term like $\arrget{S}{n}(t_0,\arridx(t_1))$ as
$t_0[t_1]$ and omit the type of $\dlfunc{set}$. The term in
\zcref[S]{example:arr} can be typeset as
$\arrsets(\arrsets(\pv{a},\arridx(\pv{i}),1),\arridx(\pv{j}),2)[\pv{i}]$.

For reading and writing to array indices, we use the functions above.
Rule $\ruleName{copyArrayIdx}$ below models reading an array \pv{a} at
index \pv{i} and storing the result in \pv{x}. We split the proof to
check for possible out of bound errors which cause a panic. We omit
the similar move version.

\[
  \seqRuleC{copyArrayIdx}
  {\sequent{\updateContext(\pv{i}<0\vee\pv{i}\ge n)}{\updateContext\ptstub{panic!()}}\\
  \sequent{\updateContext(0\le\pv{i}\wedge\pv{i}<n)}{\updateContext\upl\pv{x}\upd\pv{a}[\pv{i}]\upr\ptstub{()}}}
{\sequent{}{\updateContext\ptstub{x = a[i]}}}
{$\pv{x}: S, \pv{a}:\arrTy{S}{n}$, $S$ is \lstinline|Copy|}
\]

Similarly, the rule $\ruleName{writeArrayIdxCopy}$ handles writing
value \pv{x} to an array \pv{a} at index \pv{i}. We add an update
changing the value of \pv{a}.

\[
  \seqRuleC{writeArrayIdxCopy}
  {\sequent{\updateContext(\pv{i}<0\vee\pv{i}\ge n)}{\updateContext\ptstub{panic!()}}\\
  \sequent{\updateContext(0\le\pv{i}\wedge\pv{i}<n)}{\updateContext\upl\pv{a}\upd\arrsets(\pv{a},\arridx(\pv{i}),\mathit{se})\upr\ptstub{()}}}
{\sequent{}{\updateContext\ptstub{a[i] = $\mathit{se}$}}}
{$\pv{a}:\arrTy{S}{n}$, $S$ is \lstinline|Copy|}
\]}

References of array fields follow the same general approach as in
\zcref{sec:refs}, but are more complex, because invalid array bounds
may cause a panic which must be checked. Details are in
\diff{\ifShowApp\zcref{sec:rules-array-ref}\else\cite{DH2026TR}\fi}.

\subsection{Loop Invariants and Loop Scopes}\label{sec:loop}

Function calls and loops introduce potentially unbounded computation
and, therefore, require some form of induction principle in deductive
program verification to achieve relative completeness. Our approach is
designed to handle function calls as well as loops in their full
complexity, where treatment of the former closely follows
\cite{ABBHSU2016}. For space reasons, we discuss it only in
\diff{\ifShowApp\zcref{sec:use-contr}\else\cite{DH2026TR}\fi}.

Instead, we discuss loops in some detail: While loop invariants are a
standard concept, the rules become highly complex for loops that may
terminate due to \lstinline|break| or \lstinline|panic|. We use the
relatively recent concept of \emph{loop scope}~\cite{SW2017} to
provide concise invariant rules that handle non-standard control flow.
(For simplicity, we disregard labels, however, the presented approach
is designed to handle them.)

\begin{lstlisting}[label={lst:loop}, caption={A program computing the product of $\pv{a}$ and $\pv{b}$, storing it in $\pv{n}$. Because we use \lstinline|break| with a value, the result is also the loop's value.
},float=t]
  let mut n = 0; let old_b = b;
  loop {
    if b == 0 { break n; }
    n += a; b -= 1;
  }%\vspace{-.75cm}%
\end{lstlisting}

Consider the program in \zcref{lst:loop}, where
$\pv{a},\pv{b},\pv{old\_b}$ are program variables of
type~\lstinline|u64|. The loop is an expression and has a value: the
value of \pv{n} in the final iteration.
To safely approximate the unknown
number of loop iterations, we use a loop invariant: a \RustyDL formula
$\inv$ that holds before the first and after every subsequent
iteration.
Ignoring overflow, a sufficiently strong loop invariant $\inv$ to
prove functional correctness of the loop in \zcref{lst:loop} is the
formula (we need not strengthen the invariant with $0\le\pv{b}$, as
this is inferred from \pv{b}'s type):
\begin{equation}
  \label{eq:inv}
  \pv{n}\keyeq\pv{a}\cdot(\pv{old\_b}-\pv{b})\wedge \pv{b}\le\pv{old\_b}
\end{equation}

The \Rust compiler translates all loops to unconditional loops as in
\zcref{lst:loop}, so the invariant rules of Hoare~\cite{H1969},
Dijkstra~\cite{DS1990}, or \JavaDL~\cite{ABBHSU2016}, are unsuitable:
All make a case distinction based on the guard's value. This is
impossible here, as we are unable to extract a loop guard.
To address this and to support statements redirecting control flow,
such as \lstinline|break| and \lstinline|return|, we
use an auxiliary control statement: A \emph{loop scope}
\loopScope{x}{$p$} with \emph{index variable}~\pv{x} and body~$p$ is a
\Rust expression, where \pv{x} is a fresh program variable of type
\lstinline|bool|.

Intuitively, a loop scope represents a single generic loop iteration
and records the reason for exiting the loop in the index
variable~\pv{x}. For example, when exiting the loop through a
\lstinline|break|, \pv{x} is set to \lstinline|true|. When the loop is
not exited this way, we set \pv{x} to \lstinline|false|. Like loops
themselves, loop scopes in \RustyDL are expressions and can have a
value, namely the value of the loop they represent.
We use loop scopes in the loop invariant rule to symbolically execute
a generic loop iteration and to record how it was terminated.
Rule~$\ruleName{loopScopeInvBox}$ below matches on loops in the box
modality. It splits a proof into two branches.
\[
  \seqRule{loopScopeInvBox}{
    \sequent{}{\updateContext\inv}\\
    \sequent{}{\updateContext\mathcal{V}\Big(\inv\rightarrow\dlboxf{\loopScope{x}{$\mathit{body}$; continue;}}{\big(\\
        \quad(\pv{x}\keyeq\keybooltrue \rightarrow \dlboxf{$\pi\omega$}{\phi}) \wedge
        (\pv{x}\keyeq\keyboolfalse \rightarrow \inv)\big)}\Big)} }{
    \sequent{}{\updateContext\dlboxf{$\pi\;$loop $\mathit{body}\;\omega$}{\phi}} }
\]

The first branch ensures that $\inv$ holds when entering the loop. The
second branch puts the loop body into a loop scope with fresh index
variable $\pv{x}$ and symbolically executes the loop body once,
assuming the invariant holds. Then, the value of the index variable
$\pv{x}$ is read to determine whether the loop exits
($\pv{x}\keyeq\keybooltrue$) or further iterations are executed
($\pv{x}\keyeq\keyboolfalse$). In the first case, we must show that
postcondition $\phi$ holds, in the second that $\inv$ remains valid.

The rule executes a \emph{generic} loop iteration, where the value of
program variables occurring in $\Gamma\cup\Delta$ is not
known. Therefore, the value of such variables in $\Gamma\cup\Delta$
must be \enquote{forgotten.} This is achieved by a so-called
\emph{anonymizing update} $\mathcal{V}$ \cite{ABBHSU2016} that assigns
a fresh, unknown value to any variable written to in $\mathit{body}$.

The value of index variable $\pv{x}$ is set during symbolic execution
of the loop scope by the rules handling \lstinline|break|,
\lstinline|continue|, \lstinline|return|, etc. We illustrate the
mechanism with rules $\ruleName{breakValue}$ and $\ruleName{continue}$
and a trailing program~$p$; see also~\cite{SW2017}. Observe how
$\ruleName{breakValue}$ returns the loop scope's value
$\mathit{se}$. 
\vspace{-2ex}

\noindent\begin{minipage}[t]{.49\linewidth}
  \[
    \seqRuleW{breakValue}{
      \sequent{}{\updateContext\ptstub{\{x = true; $\mathit{se}$\}}}
    }{
      \sequent{}{\updateContext\ptstub{\loopScope{x}{break $\mathit{se}$; $p$}}}
    }
  \]
\end{minipage}\hfill
\begin{minipage}[t]{.49\linewidth}
  \[
    \seqRuleW{continue}{
      \sequent{}{\updateContext\ptstub{x = false}}
    }{
      \sequent{}{\updateContext\ptstub{\loopScope{x}{continue; $p$}}}
    }
  \]
\end{minipage}

\begin{example}
  \diff{Consider the program in \zcref{lst:loop}. Let
    $\phi\equiv\pv{n}\keyeq\pv{a}\cdot\pv{old\_b}$ and $\inv$ as
    in~\eqref{eq:inv}. After symbolically executing the assignments
    and simplifying, we have
    \[
      \seq{}{\applyUp{\pv{n}\upd
          0\parupd\pv{old\_b}\upd\pv{b}}{\dlboxf{loop \{
            ... \}}{\phi}}}.
    \]
    Applying $\ruleName{loopScopeInvBox}$ yields two sequents: first,
    the ``initially valid'' case:
    \[
      \seq{}{\applyUp{\pv{n}\upd
          0\parupd\pv{old\_b}\upd\pv{b}}{(\pv{n}\keyeq\pv{a}\cdot(\pv{old\_b}-\pv{b})\wedge\pv{b}\le\pv{old\_b})}}
    \]
    which is simplified to
    $\seq{}{0\keyeq\pv{a}\cdot(\pv{b}-\pv{b})\wedge\pv{b}\le\pv{b}}$.
    Integer simplification proves this branch.
    The second branch of $\ruleName{loopScopeInvBox}$ yields:
    \begin{align}
      \turnstyle&\applyUp{\pv{n}\upd 0\parupd\pv{old\_b}\upd\pv{b}}{\applyUp{\pv{n}\upd c_n\parupd\pv{b}\upd c_b}{\Big(\inv\rightarrow \dlboxf{\loopScope{x}{if b == 0 ...}}{}\label{eq:pres-and-use}\\
                &\quad\underbrace{\big((\pv{x}\keyeq\keybooltrue\rightarrow\dlboxf{ }{\phi}) \wedge (\pv{x}\keyeq\keyboolfalse\rightarrow\inv)\big)}_\psi\Big)}}\nonumber
    \end{align}
    Simplification and symbolically executing the if in
    \eqref{eq:pres-and-use} yields two branches
    \zcref[noname]{eq:b-zero},~\zcref[noname]{eq:b-nzero}, where
    $u\equiv\pv{old\_b}\upd\pv{b}\parupd\pv{n}\upd
    c_n\parupd\pv{b}\upd c_b$ and $\inv'\equiv\applyUp{u}{\inv}$:
    \begin{align}
      \inv',c_b\keyeq 0&\turnstyle\applyUp{u}{\dlboxf{\loopScope{x}{break n; n+=a; b-=1; continue;}}{\psi}} \label{eq:b-zero} \\
      \inv',c_b\not\keyeq 0&\turnstyle \applyUp{u}{\dlboxf{\loopScope{x}{n+=a; b-=1; continue;}}{\psi}} \label{eq:b-nzero}
    \end{align}
    Applying $\ruleName{breakValue}$ to \zcref[noname]{eq:b-zero}
    drops the remaining loop body and the loop scope:
    \begin{align}
      \inv',c_b\keyeq 0\turnstyle\applyUp{u}{\dlboxf{\{x = true; n\}}{\psi}}
    \end{align}
    Hence, we know \pv{x} is true in this branch, allowing us to
    simplify the sequent to
    \begin{align}
      \inv',c_b\keyeq 0\turnstyle\applyUp{u}{(\pv{n}\keyeq\pv{a}\cdot\pv{old\_b})}
    \end{align}
    Applying the updates yields \zcref[noname]{eq:inv-impl-post},
    easily proven by arithmetic simplification.
    %
    \begin{align}
      c_n\keyeq\pv{a}\cdot(\pv{b}-c_b)\wedge c_b\le\pv{b},c_b\keyeq 0&\turnstyle c_n\keyeq\pv{a}\cdot\pv{b} \label{eq:inv-impl-post}
    \end{align}
    The proof steps for \zcref[noname]{eq:b-nzero} follow similar
    lines.}
\end{example}

\subsection{Implementation of Calculus Rules}
\label{sec:calc-impl}%
%
We implemented \RustyDL and its calculus based on the \KeY
tool,\footnote{See implementation and examples at
  \url{https://github.com/Drodt/key/tree/rusty}.}  leveraging the
general data structures and rule matching algorithm of that
system. The proof system is not foundational; we rely on theories of
integers, sequences, etc.

All rules are implemented in a domain specific language for writing
schematic program logic rules in a sequent calculus called
\enquote{taclets}~\cite{BBDHLPUW2025}. They can be reviewed, checked,
and partially, even mechanically proven independently from the
remaining system. There has also been work to prove the rules sound
relative to program semantics and to validate the transformation
rules~\cite{ARS2005,T2005}.

The \KeY system is time-proven. Thus, we argue, the core is
trustworthy and the implementation faithfully reflects the calculus
rules above.
At the moment, the implementation is a prototype able to manually and
automatically prove a number of examples showcasing several \Rust
features, in particular, borrowing, loops, arrays, tuples, and (some)
enums. Most instructive is a verification of binary search, creating a
proof of $4260$ rule applications in $2.1$ seconds.

\section{Related Work}\label{sec:related}

Our approach to deductive verification was inspired by
\KeY~\cite{ABBHSU2016}.
Several other tools have been developed in the last years to verify \Rust
programs. 
%
\emph{Creusot}~\cite{DJM2022} translates safe \Rust programs to the
intermediate language and verification tool \emph{Why3}~\cite{FP2013}
which generates verification conditions that are discharged with SMT
solvers. Mutable references are modeled using
prophecies~\cite{JLPRTDJ2020}, in effect writing the value of the
mutable reference back to the lender once the reference is dropped.
The \emph{Prusti} tool~\cite{ABFGMMPS2022} is similar to Creusot, but
based on separation logic, and translates to
\emph{Viper}~\cite{MSS2016}. \emph{Aeneas}~\cite{HP2022} translates
safe \Rust to several functional back ends, removing mutable
references entirely. All these tools rely on translation to
intermediate languages and allow no interaction, whereas our dynamic
logic permits a HIL interaction pattern.

\emph{Verus}~\cite{LHCBSZHPH2023} enables verification of unsafe
\Rust, based on verification condition generation and SMT
solving. Presently, it is unclear how \RustyDL might be extended to
unsafe \Rust, the \enquote{hybrid} approach of~\cite{ADMG2024} being
strong contender.

An alternative to logic modeling of references are \emph{permissions}
\cite{Boyland13} to deal with read and read-and-write permissions to
memory locations. Permissions were implemented in the \KeY tool
in~\cite{HM2015}, but here we chose mutable updates, because they
cause less overhead and easily extend the existing update mechanism.

\section{Conclusion and Future Work}\label{sec:concl}

We presented \RustyDL, a dynamic logic for a subset of \Rust, as well
as a sequent calculus to prove validity of \RustyDL formulas. We gave
solutions for logic modeling of crucial \Rust features with reasonable
complexity. This demonstrates that the \KeY approach to HIL program
verification can be applied to \Rust.

The \RustyDL calculus presented here was implemented on top of the
\KeY verification tool as a proof-of-concept.
We used that implementation to successfully verify several \Rust
functions, including ones containing loops and references, however, a
fuller evaluation needs to wait until the degree of automation is
sufficiently high.

Accordingly, future work can be divided into two directions: First,
extending the subset of \Rust to include Traits, iterators, pattern
matching, and generic functions. This is time consuming, yet
straightforward. Verification of unsafe \Rust also should be
considered, but one has to carefully weigh to which extent. It is
likely that unsafe \Rust requires a more explicit memory model than we
introduced here; still, one would like this model to be minimal and
refer to the safe version where possible. A thorough investigation
remains to be done. 
Second, we are improving the implementation: flesh out the
specification language, similar to JML~\cite{JML-Ref-Manual} and
implement a graphical user interface.


\bibliographystyle{splncs04}
\bibliography{RustyDL}

\ifShowApp
\appendix

\renewcommand{\theHsection}{A\arabic{section}}

\section{Calculus Rules}

\subsection{Update Simplification Rules}\label{app:upd-rules}

See \zcref{tab:upd-simp}. 

\begin{table}[t]
  \centering
  \begin{tabular}{rclr}
    $\applyUp{u_1\parupd\pv{a}\upd t_1\parupd u_2\parupd\pv{a}\upd t_2\parupd u_3}{t}$ & $\leadsto$ & $\applyUp{u_1\parupd u_2\parupd\pv{a}\upd t_2\parupd u_3}{t}$ & \\ 
                                                                                       & & where $t\in\DLTrm{A}\cup\DLFor\cup\Updates$ & \\
    $\applyUp{u_1\parupd\pv{a}\upd t'\parupd u_2}{t}$ & $\leadsto$ & $\applyUp{u_1\parupd u_2}{t}$ & \\ 
                                                                                       & & where $t\in\DLTrm{A}\cup\DLFor\cup\Updates, \pv{a}\not\in\freeProgVars(t)$ & \\
    $\applyUp{u_1}{\applyUp{u_2}{t}}$ & $\leadsto$ & $\applyUp{u_1\parupd\applyUp{u_1}{u_2}}{t}$ & \\ 
                                                                                       & & where $t\in\DLTrm{A}\cup\DLFor\cup\Updates$ & \\
    $\applyUp{u\parupd\skipUp}{t}$ & $\leadsto$ & $\applyUp{u}{t}$ where $t\in\DLTrm{A}\cup\DLFor\cup\Updates$ & \\ 
    $\applyUp{\skipUp\parupd u}{t}$ & $\leadsto$ & $\applyUp{u}{t}$ where $t\in\DLTrm{A}\cup\DLFor\cup\Updates$ & \\ 
    $\applyUp{\skipUp}{t}$ & $\leadsto$ & $t$ where $t\in\DLTrm{A}\cup\DLFor\cup\Updates$ & \\ 
    $\applyUp{u}{v}$ &  $\leadsto$ & $v$ where $v\in\VSym$ & \\ 
    $\applyUp{u}{f(t_1,\ldots,t_n)}$ &  $\leadsto$ & $f(\applyUp{u}{t_1},\ldots,\applyUp{u}{t_n})$ where $f\in\FSym\cup\PSym$ & \\ 
    $\applyUp{u}{(\condTerm{\phi}{t_1}{t_2})}$ &  $\leadsto$ & $\condTerm{\applyUp{u}{\phi}}{\applyUp{u}{t_1}}{\applyUp{u}{t_2}}$ & \\ 
    $\applyUp{u}{\neg\phi}$ &  $\leadsto$ & $\neg\applyUp{u}{\phi}$ & \\ 
    $\applyUp{u}{(\phi\bullet\psi)}$ &  $\leadsto$ & $\applyUp{u}{\phi}\bullet\applyUp{u}{\psi}$ where $\bullet\in\{\wedge,\vee,\rightarrow,\leftrightarrow\}$ & \\ 
    $\applyUp{u}{\keyquantification{\mathcal{Q}}{A}{v}\phi}$ &  $\leadsto$ & $\keyquantification{\mathcal{Q}}{A}{v}\applyUp{u}{\phi}$ where $\mathcal{Q}\in\{\forall,\exists\},v\not\in\freeVars(u)$ & \\ 
    $\applyUp{u}{(\pv{a}\upd t)}$ &  $\leadsto$ & $\pv{a}\upd\applyUp{u}{t}$ & \\ 
    $\applyUp{u}{(t_1\dupd t_2)}$ &  $\leadsto$ & $\applyUp{u}{t_1}\dupd\applyUp{u}{t_2}$ & \\ 
    $\applyUp{u}{(u_1\parupd u_2)}$ &  $\leadsto$ & $\applyUp{u}{u_1}\parupd\applyUp{u}{u_2}$ & \\ 
    $\refM{\place{\pv{a}}}\dupd t$ &  $\leadsto$ & $\pv{a}\upd t$ & \\ 
    $\refM{\arrplace{\refM{\place{\pv{a}}}}{t_1}}\dupd t_2$ & $\leadsto$ & $\pv{a}\upd\arrsets(\pv{a},t_1,t_2)$  & \\
    $\applyUp{\pv{a}\upd t}{\pv{a}}$ &  $\leadsto$ & $t$ & \\[1ex] 
  \end{tabular}
  \caption{Simplification rules for updates}\label{tab:upd-simp}
\end{table}

\subsection{Rules for Array References}
\label{sec:rules-array-ref}

Consider rule $\ruleName{borrowMutIdxArr}$: We want to get a mutable
reference to index \pv{i} of the array referenced by \pv{a}. Again, we
perform a bounds check. We update \pv{x} to a mutable reference of
place $\pv{a}[\pv{i}]$.
\[
  \seqRuleWC{borrowMutIdxArr}
  {\sequent{\updateContext(\pv{i}<0\vee\pv{i}\ge n)}{\updateContext\ptstub{panic!()}}\\
  \sequent{\updateContext(0\le\pv{i}<n)}{\updateContext\applyUp{\pv{x}\upd\refM{\arrplace{\pv{a}}{\pv{i}}}}{\ptstub{()}}}}
  {\sequent{}{\updateContext\ptstub{x = \&mut a[i]}}}
  {$\pv{a}:\RefM{\arrTy{A}{n}}, \pv{x}:\RefM{A}, \pv{i}:\Int$}
\]

\begin{example}
  Let
  $\pv{a}:\arrTy{\Int}{4},\pv{x}:\RefM{\arrTy{\Int}{4}},\pv{y}:\RefM{\Int}$
  be program variables. We start in the sequent
  \[
    \seq{}{\dlf{x = \&mut a; y = \&mut x[0]; *y = 8;}{\phi}}\enspace.
  \]
  
  Full symbolic execution of the program fragment yields the sequent
  \[
    \seq{}{\applyUp{\pv{x}\upd\underbrace{\refM{\place{\pv{a}}}}_{r_1}}{\applyUp{\pv{y}\upd\refM{\arrplace{\pv{x}}{0}}}{\applyUp{\pv{y}\dupd 8}{\phi}}}}\enspace.
  \]
  
  Simplifying the first two updates gives us the following
  sequent. Observe that we use the mutable reference stored in \pv{x}
  within \pv{y}'s value.
  \[
    \seq{}{\applyUp{\underbrace{\pv{x}\upd r_1\parupd\pv{y}\upd\refM{\arrplace{r_1}{0}}}_{u_1}}{\applyUp{\pv{y}\dupd 8}{\phi}}}
  \]

  Combining the updates yields the following sequent:
  \[
    \seq{}{\applyUp{u_1\parupd\refM{\arrplace{r_1}{0}}\dupd 8}{\phi}}
  \]

  Next, we resolve the mutating update using
  $\ruleName{mutatingArrToElementary}$:
  \begin{align*}
    \seq{}{\applyUp{&u_1\parupd\pv{a}\upd\arrsets(\pv{a},\arridx(0),8)}{\phi}}\enspace.
  \end{align*}
\end{example}

\section{Applying Function Contracts}\label{sec:use-contr}%


Consider a function call \lstinline|r = f(p$_1$, $\ldots$, p$_n$)|,
where \pv{f} is a function with arguments
\lstinline|a$_1$: T$_1$, $\ldots$, a$_n$: T$_n$| and
$\pv{p}_1,\ldots,\pv{p}_n,\pv{r}$ are program variables of compatible
types. There are two ways of resolving such a call in \RustyDL.

First, we can \enquote{inline} the body of function \pv{f}, creating
local variables for parameters, syntactically replacing return
expressions with assignments to \pv{r}, etc. This approach cannot
handle unbounded recursion, leads to repeated symbolic execution of
the same \Rust code and, therefore, often results in state explosion.

Second, we can use a suitable declarative abstraction to reason about
\rust{f}'s behavior without symbolically executing its body. To
accomplish this, \zcref{def:contract} introduces the notion of a
function's \emph{contract}.

\begin{definition}[Function Contract]\label{def:contract}
  Let $\Crate$ be a valid Rust crate and let
  \lstinline|fn f(a$_1$: T$_1$, $\ldots$, a$_n$: T$_n$) -> T| be a
  function in $\Crate$. A triple $(\pre,\post,\myterm)$ is a
  \emph{contract} for \lstinline|f|, consisting of:
  \begin{enumerate*}[(1)]
  \item formula $\pre$ as precondition,
  \item formula $\post$ as postcondition,
  \item an optional term $\myterm$ for the termination witness.
  \end{enumerate*}
  These components have access to the arguments of \rust{f},
  $\pv{a}_1:\rust{T}_1,\ldots,\pv{a}_n:\rust{T}_n$, and the
  postcondition in addition has access to the result of \rust{f},
  $\pv{res}: \rust{T}$.
\end{definition}

The rule $\ruleName{useFnContract}$ in \zcref{fig:use-fn} replaces a
function call with the function's contract, assuming that this
contract is correct, which must be proven separately.

\begin{figure}[t]
  {\centering
  \[
    \seqRule{useFnContract}{
      \sequent{}{\updateContext\mathcal{V}(\pre\wedge\mathit{paramsInRange}\wedge\myterm\prec\mby)}\\
      \sequent{}{\updateContext\mathcal{V}\mathcal{W}(\post\wedge\mathit{inRange}_\rust{T}(\pv{res})\rightarrow\ptstub{r = res})}
    }{
      \sequent{}{\updateContext\ptstub{r = f($\mathit{se}_1$, $\ldots$, $\mathit{se}_n$)}}
    }
  \]
} Where
\begin{itemize}
\item \lstinline|fn f(a$_1$: T$_1$, $\ldots$, a$_n$: T$_n$) -> T| is a
  function,
\item $(\pre,\post,\myterm)$ is a contract for \rust{f},
\item
  $\mathcal{V}:=\upl\pv{a}_1\upd\mathit{se}_1\parupd\cdots\parupd\pv{a}_n\upd\mathit{se}_n\upr\in\Updates$
  initializes the parameters,
\item $\mathcal{W}:=\upl\pv{b}_1\dupd c_1\parupd\cdots\parupd\pv{b}_m\dupd
c_m\upr$ is an anonymizing update,
\item $\pv{b}_1:\RefM{A_1},\ldots,\pv{b}_m:\RefM{A_m}$ are exactly the
  parameters of \rust{f} with mutable reference type,
\item $c_1,\ldots,c_m$
are fresh constants,
\item and $\mby$ is a termination witness if this is a recursive call.
\end{itemize}
  \caption{Function contract rule}
  \label{fig:use-fn}
\end{figure}

The rule has two premises: The first establishes that the function's
precondition is satisfied in the current symbolic state, after
$\mathcal{V}$ updated the formal parameters to the actual
parameters. Additionally, we show that the termination witness is
strictly decreasing, with $\prec$ a well-founded binary relation over
$\myterm$'s type. This is necessary for recursive calls: If $\myterm$
strictly decreases with each recursive call, well-foundedness ensures
that recursion is bounded and the call terminates.

The second premise allows us to assume the function's postcondition
and continue with the proof. However, the called function may change
parts of the state if we pass a mutable reference. Therefore, we add
the \emph{anonymizing update} $\mathcal{W}$, which assigns fresh
constants to all possibly mutated places, see \zcref{sec:loop} for
further explanations.

\begin{example}
  Consider the recursive function
  \lstinline|fn mul(a: u64, b: u64, c: &mut u64)| with contract
  $(\keytrue, \derefM(\pv{c})\keyeq\pv{a}\cdot\pv{b},\text{---})$. It
  computes the result of multiplying \pv{a} times \pv{b} and writes
  the result to \pv{c}.

  Assume the goal sequent has the following shape with program
  variables \pv{x}, \pv{y} of type \lstinline|u64|, \pv{z} of type
  \lstinline|&mut u64|, and \pv{r} of type \lstinline|()|:
  \[
    \seq{\typerestrict{\rust{u64}}{\pv{x}},
      \typerestrict{\rust{u64}}{\pv{y}}, \typerestrict{\rust{\&mut
          u64}}{\pv{z}}}{\dlf{r = mul(x, y, z);}{\phi}}
  \]
  
  Applying $\ruleName{fnUseContract}$ yields as first premise the
  sequent:
  \begin{align*}
    &\typerestrict{\rust{u64}}{\pv{x}},
      \typerestrict{\rust{u64}}{\pv{y}}, \typerestrict{\rust{\&mut
          u64}}{\pv{z}}\\
    \seq{}{&\applyUp{\underbrace{\pv{a}\upd\pv{x}\parupd\pv{b}\upd\pv{y}\parupd\pv{c}\upd\pv{z}}_{\mathcal{V}}}{(\keytrue\wedge\paramsInRange)}}
  \end{align*}
  
  This can easily be discharged, because all parameters are in range
  of their respective types. The second sequent looks, slightly
  simplified, as follows:
  \begin{align*}
    &\typerestrict{\rust{u64}}{\pv{x}},
      \typerestrict{\rust{u64}}{\pv{y}}, \typerestrict{\rust{\&mut
          u64}}{\pv{z}}\\
    \seq{}{&\mathcal{V}\applyUp{\pv{c}\dupd d}{(\derefM(\pv{c})\keyeq\pv{a}\cdot\pv{b}\rightarrow\dlf{r = res;}{\phi})}}
  \end{align*}
  where $d$ is a fresh constant of sort $\Int$, anonymizing \pv{c}'s
  lender.

  The fact that \pv{c} contains the product of \pv{a} and \pv{b} can
  now be used in the subsequent proof of $\phi$.
\end{example}

\fi

\end{document}